\documentstyle[11pt,epsfig]{article}

\def\thefootnote{\fnsymbol{footnote}}
\def\roughly#1{\mathrel{\raise.3ex\hbox{$#1$\kern-.75em%
\lower1ex\hbox{$\sim$}}}}

\def\lsim{\roughly<}
\def\gsim{\roughly>}
\def\be{\begin{eqnarray}}
\def\ee{\end{eqnarray}}

\def\ben{\begin{enumerate}}
\def\een{\end{enumerate}}
\def\beitem{\begin{itemize}}
\def\eitem{\end{itemize}}

\def\thefootnote{\fnsymbol{footnote}}
\newcommand{\beq}{\begin{eqnarray}}
\newcommand{\eeq}{\end{eqnarray}}

\def\bi{\begin{itemize}}
\def\ei{\end{itemize}}

\def\del{\partial}
\def\L{{\cal L}}

\long\def\beginomit#1\endomit{}
\def\np{{Nucl. Phys.}}
\def\prl{Phys. Rev. Lett.}
\def\pr {Phys. Rev.}

\def\pl{Phys. Lett.}
\def\M{{\cal M}}

\def\sufxsuf{$SU(N_f)\times SU(N_f)$}
\def\su3xsus3{$SU(3)\times SU(3)$}
\def\su2xsu2{$SU(2)\times SU(2)$}

\def\su2{$SU(2)$}
\def\su3{$SU(3)$}
\def\onebod{j^{(1)}}

\begin{document}


\begin{titlepage}\begin{center}

\hfill{SNUTP-95-081}

\hfill{nucl-th/9508046}

\hfill{August 1995}
\vskip 0.4in
{\Large\bf THE CHANGES IN PROPERTIES OF HADRONS IN NUCLEI AND NUCLEAR
MATTER}\footnote{Invited talk given at ``International Nuclear Physics
Conference," August 21-26, 1995, Beijing, China.}
\vskip 1.2in
{\large  Mannque Rho}\\
\vskip 0.1in
{\large \it Service de Physique Th\'{e}orique, CEA  Saclay}\\
{\large\it 91191 Gif-sur-Yvette Cedex, France}\\
{\large and}\\
{\large\it Center for Theoretical Physics}\\
{\large\it Seoul National University}\\
{\large\it Seoul 151-742, Korea}
\vskip .6in
\vskip .6in

{\bf ABSTRACT}\\ \vskip 0.1in
\begin{quotation}
\noindent
The changes of hadron properties in dense and/or hot matter are
discussed in terms of effective chiral Lagrangians with the
parameters of the theory scaled in a simple way. The phenomenologically
successful Walecka model is identified as a mean field chiral Lagrangian
with the scaled parameters. Kaon condensation and chiral restoration
transitions can be described within the same mean field framework.
\end{quotation}
\end{center}\end{titlepage}


\renewcommand{\thefootnote}{\#\arabic{footnote}}


\indent

Quantum chromodynamics (QCD) tells us that most, if not all,
of light hadron masses are
generated spontaneously by the breaking of chiral symmetry from
{\sufxsuf} to diagonal $SU(N_f)$ where $N_f$ is the number of flavors,
equal to 2 without strangeness and 3 with. It is also widely believed that
as a hadronic system is heated to high temperature or compressed
to high density, the broken symmetry will get restored in a way paralleling
what happens in condensed matter physics.
A natural consequence of the restoration of the chiral symmetry must then
be that the spontaneously generated masses disappear as density
(and perhaps also temperature) is increased. The question we are
raising is how does this ``shedding of mass" occur?

This question is at the core of the fundamental theory of matter:
How is the mass generated, starting with the lightest object like neutrinos
to the heaviest detected particle like the top quark?

The aim of this talk is to describe how hadron properties get modified
in medium as the
system is heated or compressed. That is, immerse a hadron in medium and
compress the system or heat it. What does one expect to see happening?

To answer this question, let me start with the simplest nuclear system,
namely the deuterium. Let us look at what happens when a soft photon is
sent in to probe the system. Consider therefore the well-known inverse
process
\be
n + p\rightarrow d +\gamma\label{npcap}
\ee
at thermal neutron energy. This process was first explained in a
quantitative way by Riska and Brown \cite{riskabrown} in 1973. What
I will do here is to describe it more accurately in a modern QCD framework.
Since it is a very low-energy process, QCD can be
represented by an effective chiral Lagrangian field theory. This is because
at long wavelength limit, chiral perturbation theory (ChPT) is believed to be
exactly equivalent to QCD \cite{chptqcd}. This invites us to attempt to
describe
(\ref{npcap}) in terms of a chiral Lagrangian. There is one basic problem
in doing this and that has to do with the description of the deuteron
in QCD: we do not really know how to derive the deuteron starting {\it
directly}
from QCD. Since we are focusing on the chiral aspect of the problem, however,
the solution might be sought in a chiral Lagrangian approach to the
structure of the deuteron. If the number of colors $N_c$ is in some sense
big -- which gives rise to what is known as the ``large $N_c$ limit" --
then the effective Lagrangian is given by meson fields only as we know
from the skyrmion structure. There is some important progress in
obtaining a bound deuteron, recently through the work of Manton \cite{manton},
as a baryon number 2 skyrmion, but we are still far  from
understanding it quantitatively. However there is an indirect approach to this
which is consistent with QCD and which has the potential to be quantitatively
accurate, namely that as we have known all along, the deuteron is made up of
a proton and a neutron bound by meson exchanges: In the framework of QCD,
the nucleon may be gotten from a large $N_c$ Lagrangian as a soliton
(skyrmion) but this is now known to be
equivalent, at least in the large $N_c$ limit,
to having the nucleon as a matter field
in the chiral Lagrangian. We are thus led to consider a chiral Lagrangian
that contains baryons (nucleon, $\Delta$ etc.), pseudo-Goldstone bosons $\pi^i$
(pions, kaons etc.), vectors $V_\mu=
\omega_\mu, \rho_\mu, \cdots$ etc. with suitable chiral invariant couplings.
There have been some attempts to compute the
deuteron from such a Lagrangian in chiral perturbation theory
\cite{vankolck} but at present the calculation can be done only at low orders
since higher order calculations would involve too many parameters to be
completely determined from available experiments. Luckily for our purpose,
we need not compute the deuteron from first principles as I shall
argue below.

The important point to note
is that certain aspects of the deuteron which have to do
with the chiral symmetry structure of hadrons can be probed by
the process (\ref{npcap}) {\it without knowing}
how to get the nucleus itself from
a chiral perturbation theory, as recently discussed by Park, Min and Rho
\cite{pmr}. Briefly the argument goes as follows:
For physics with energy scale much less than the chiral scale $\Lambda_\chi
\sim 4\pi f_\pi\sim 1$ GeV (where $f_\pi$ is the pion decay constant
$\sim 93$ MeV), the relevant Lagrangian is, schematically,
\be
\L=\sum_i \L_i [B, U, \M]
\ee
where $B_i$ are the baryon fields (both octet and decuplet), $U$ the unitary
pseudo-Goldstone fields $U=e^{i\pi/f_\pi}$ and $\M$ the quark mass matrix.
The degrees of freedom more massive than the chiral scale $\Lambda_\chi$ are
integrated out, appearing implicitly in the counter terms of the Lagrangian.
Chiral symmetry requires that there be only derivative couplings apart from
terms involving the
mass matrix and hence effectively the Lagrangian is an
expansion in $\del/\Lambda_\chi$ and $\M/\Lambda_\chi^2$. Since the baryon
mass is $\sim \Lambda_\chi$, the baryons should be introduced as static
matter fields, so the derivative on the baryon field does not involve
time derivatives. ChPT is then a systematic perturbative expansion
in powers of $Q$, say, $Q^n$, where $Q$ is the energy-momentum scale
being probed,
with suitable counter terms to remove divergences and to take account of the
degrees of freedom that are
integrated out. Now in applying this theory to nuclear
systems, we need to separate the class of Feynman diagrams into two,
one ``irreducible" and the other ``reducible." It is in calculating the
irreducible diagrams that ChPT enters. The reducible diagrams -- that
cannot be treated by ChPT because of infrared singularity -- are incorporated
by solving a Lippman-Schwinger equation or Schr\"odinger equation with the
potential obtained with the irreducible graphs by ChPT. This is how
bound states are to be treated in ChPT. Now in calculating the process
(\ref{npcap}), we can write the EM current in two terms, one the
single-particle
current $j^{(1)}$ and the other the two-body current $j^{(2)}$
\be
j_\mu^{EM}=j_\mu^{(1)} + j_\mu^{(2)}.\label{current}
\ee
The former
is called ``impulse approximation current" and the latter ``exchange
current." For the system considered, we terminate with the two-body current.
Later we will see that in heavy nuclei there can enter many-body currents,
some of which become quite important.

Very accurate wave functions for the final deuteron and the initial
neutron-proton system obtained from some accurate phenomenological potential
such as the Argonne $v_{18}$ potential \cite{v18}
(or a potential calculated in a high order ChPT if it is feasible)
would correspond to a high order chiral expansion since
the Schr\"odinger equation sums a certain class of
chiral series to all orders and
presumably the phenomenological potential also subsumes all orders of chiral
perturbation. Now the idea is to compute the matrix element of the
current (\ref{current}) in chiral perturbation expansion in such a way
that is consistent with the calculation of the wave functions.
The calculation of the one-body current $\onebod$ is without ambiguity.
That of the two-body exchange current is somewhat subtle, requiring
a careful sorting of irreducible and reducible contributions such that
the reducible ones are suitably accounted for in the one-body term
with the accurate wave functions. In this way, the Ward identities
associated with the conserved vector current are satisfied to a
given chiral order in the EM current. This is a numerically accurate
procedure.

\begin{figure}
\centerline{\epsfig{file=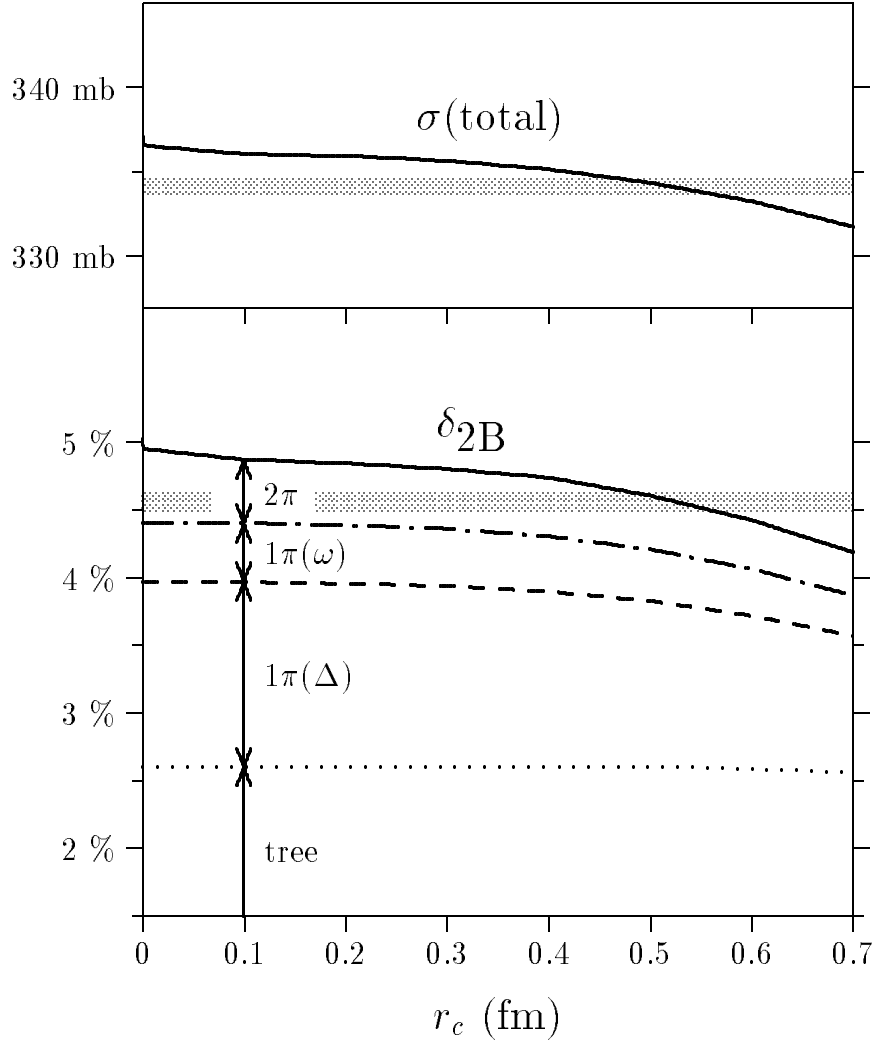}}
\caption[np]{The np capture rate calculated in chiral perturbation theory.
The predicted capture rate (upper plot) $\sigma=(334\pm 3)$ mb agrees with
the experimental value $(334.2\pm 0.5)$ mb. The lower plot shows
the ratios of the matrix elements of the two-body terms over the one-body term.
``Tree" corresponds to the leading chiral order one-pion exchange term (with
the blob in Figs.\ref{graphs}(a,b) replaced by a bare vertex),
``$1\pi (\omega)$" and ``$1\pi (\Delta)$" correspond to the next-to-leading
order corrections involving the $\omega$ meson and $\Delta$ resonance,
respectively, to the one-pion exchange tree terms Figs.\ref{graphs}(a,b).
``$2\pi$"
is the genuine loop correction to the tree contribution. The hadrons appearing
in this calculation all have free-space properties. }\label{nprates}
\end{figure}

A recent calculation \cite{pmr} of the process (\ref{npcap}) is given in
Fig.\ref{nprates}. The calculation was done to order $Q^3$ which corresponds
to next-to-next-to leading order in chiral expansion, that is, to one-loop
order.   Given the nucleon mass $m_N$, the pion decay constant
$f_\pi$, the pion mass $m_\pi$, the axial coupling constant $g_A$
and the vector meson mass $m_V$ all {\it determined in free space},
all the parameters that appear in the theory
are fixed in the theory except for
the hard core radius $r_c$ in the wave function reflecting on our inability
to handle in ChPT
very short-range physics. The remarkable agreement with
the experiment shows that the two-nucleon systems we are looking at
are made up of two nucleons with their properties as given in free space:
the chiral Lagrangian with the vacuum values of hadron parameters
describes nature remarkably well at low chiral orders.

\begin{figure}
\centerline{\epsfig{file=pmr3fig1.eps}}
\caption{Generic graphs contributing to exchange currents.
(a) and (b) are one-pion exchange and (c) represents multipion and/or
heavy-meson exchange currents. The large filled circles represent one-nucleon,
one-pion irreducible graphs, the solid line the nucleon, the dotted line the
pion and the wiggly line the current $j_\mu$.}\label{graphs}
\end{figure}

A very valuable information for heavier and denser nuclei
is lodged in some of the terms that are
{\it negligible} in the process (\ref{npcap}). As I showed elsewhere
\cite{mr91},
two-body currents involving four baryon fields (e.g. Fig.\ref{graphs}(c)
with the blob replaced by a point) like
\be
\bar{B}\Gamma_\mu B \bar{B}\tilde{\Gamma} B\label{fourbaryon}
\ee
with $\tilde{\Gamma}$ and $\Gamma_\mu$ representing some Lorentz scalar
and Lorentz vector quantities consistent with chiral symmetry, respectively,
are subdominant in the chiral counting and can be ignored. Now
terms like (\ref{fourbaryon}) show up in the chiral Lagrangian
as a result of integrating out heavy degrees of freedom with an energy scale
$E\gsim \Lambda_\chi$. Consider for instance the scalar meson $\sigma$
that plays an important role in effective field theory of nuclei, a prototype
of which being the
Walecka model. In free space, there is no low-lying scalar that can appear
in the low-energy chiral Lagrangian. But there is a high-lying
scalar field that can be associated with the trace anomaly of QCD
\be
(T_\mu^\mu)_{QCD}=-\frac{\beta (g)}{2g} (G_{\mu\nu}^a)^2\sim \chi^4
\ee
where $\beta$ is the QCD beta function, $g$ the color gauge coupling constant,
$G_{\mu\nu}^a$ the gluon field tensor and $\chi$ the scalar glueball field.
Now the $\chi$ field is massive, with $m_\chi\sim 2$ GeV, so this
degree of freedom appears only in the counter terms. It will give rise to
a term like
\be
\kappa \bar{B}j_\mu^{(1)}B \bar{B}B\label{sigmacurrent}
\ee
with the coefficient $\kappa$ suppressed by the power $(Q/m_\chi)^2$.
In the process (\ref{npcap}), $Q$ is of order of 40 MeV, so $(Q/m_\chi)^2
< 10^{-3}$. Stated differently an effective two-body term like
(\ref{sigmacurrent}) will be screened by the short-range correlation implicit
in the wave functions.

The situation is quite different, however, in dense nuclear medium.
As discussed in \cite{brpr}, as density increases, the scalar
(or precisely the quarkish component of the scalar)
$\chi$ moves downwards in energy and at some high density, it joins
the triplet of nearly massless pions to make up the quartet of the $O(4)$
symmetry of chiral symmetry.
The merging presumably takes place at the chiral transition point
discussed below.
The point is that as discussed by Beane and van Kolck
\cite{beane}, in order to reconcile Weinberg's
``mended symmetry" \cite{mendedwein}
with effective chiral Lagrangians at some shorter length scale, the scalar
field must come down as a dilaton. Now if it comes down below
the chiral scale $\Lambda_\chi$, then we can no longer consider the scalar as a
counter-term contribution. It will strongly couple to low-mass multipion
excitations giving among others
what is usually taken as a scalar field $\sigma$ in
effective nuclear forces \cite{brpr}. All other hadrons (other than Goldstone
bosons) will also couple to this
scalar as well and will undergo a mass shift as density increases.
A chiral Lagrangian that accounts for this phenomenon has been shown to
lead to the Brown-Rho scaling \cite{br91}
\be
m_B^\star/m_B\approx m_V^\star/m_V\approx m_\sigma^\star/m_\sigma
\approx f_\pi^\star/f_\pi\approx \cdots\label{BR}
\ee
where $B$ stands for baryons, $V$ for vector mesons and $\sigma$ for the
dilatonic scalar. The star stands for density-dependent quantities.
Such a chiral Lagrangian effective in dense system will then contain
these effective
constants instead of the free-space values used in the process
(\ref{npcap}) while preserving the free-space chiral symmetry. As I will argue,
the consequence of this scaling can be significant in heavy nuclei and nuclear
matter.

One can already
see the effect of this scaling in finite nuclei. One clear case is
the axial charge transition in nuclei. It was shown in \cite{pmr,pmr93} that to
order $Q^3$ in chiral expansion the axial charge transition matrix element
in heavy nuclei
\be
A (0^\pm)\rightarrow B(0^\mp),\ \ \ \ \Delta T=1\label{axial}
\ee
is enhanced with respect to the impulse approximation
by the factor
\be
\epsilon_{\tiny MEC}= \frac{m_N}{m_N^\star} (1+ R)
\ee
where $R$ is the ratio of the exchange current matrix element Fig.\ref{graphs}
to the impulse approximation calculated
with matter-free-space constants. In (\ref{axial}), $R$ is
essentially given by Fig.\ref{graphs}(a) with the bare coupling and
is given by $R\approx 0.5$ with a small variation with density of the
system. Now at nuclear matter density, $m_N^\star/m_N\approx 0.75$, so
\be
\epsilon_{\tiny MEC}\approx 2.
\ee
In light nuclei, we expect $\epsilon_{\tiny MEC}\gsim 1.5$.
These results are in agreement what was found experimentally \cite{warburton}.

One predicts a similar effect in magnetic moments of heavy nuclei but
here one has to include other effects of equal
importance present in the vector current case.
For instance, the ``back-flow" correction due to Galilean
invariance cancels almost completely the corrections coming from
the scaled nucleon mass.

We now turn to an important issue of making a bridge
between the chiral theory and Walecka mean field theory of nuclei
and nuclear matter which is found to be very successful. I wish
to show here, following \cite{br95-1},
that Walecka theory is equivalent to the chiral Lagrangian theory
at mean field {\it with} the BR scaling and that
this would allow a treatment of fluctuations into different
flavor directions (such as strangeness) in a way consistent with
the properties of normal nuclear matter. To do this we can focus on
the four-fermi interactions allowed in chiral Lagrangians that are
relevant in making contact with Walecka theory,
\be
\L_{4f}=\alpha \left(\bar{B}B\right)^2 +\beta \left(\bar{B} v^\mu B\right)^2
\label{4flag}
\ee
where $\alpha$ and $\beta$ are dimension -2 constants and we are using
the heavy-fermion formalism so that $v_\mu$ is the velocity four-vector
of the heavy baryon. Let us imagine that the first term of (\ref{4flag})
arises from integrating out the heavy chiral singlet scalar $\chi$ and
the second term from integrating out a heavy chiral singlet vector meson
$\omega$. We can include other degrees of freedom in a similar way but
we will not need them for symmetric nuclear matter that we shall consider.
In this case, we can identify the constants
\be
\alpha=\frac{g_\chi^2}{2m_\chi^2}, \ \ \ \
\beta=-\frac{g_\omega^2}{2m_\omega^2}.
\ee
We now ask what happens to this Lagrangian when it is immersed in dense and/or
hot matter. In mean field, we get the nucleon scalar potential $S_N$ and
vector potential $V_N$ as
\be
S_N &=& -\frac{{g_\sigma^\star}^2}{{m_\sigma^\star}^2} \rho_s,\label{SN}\\
V_N &=& \frac{9}{{8f_\pi^\star}^2} \rho\label{VN}
\ee
where $\rho_s$ is the scalar density and $\rho$ the vector density.
In obtaining (\ref{VN}), we have used $SU(3)_f$ relations together with
KSRF relation which is known to hold well and put the stars in (\ref{SN})
and (\ref{VN}) to indicate
that they are in-medium quantities. Now comparing with the phenomenology
with Walecka model, we find that the identification requires that
(\ref{BR}) holds with
\be
\frac{f_\pi^\star}{f_\pi}\approx 0.77.
\ee
Together with what we found in the case of the axial charge transition,
we come to the conclusion that the mean field Walecka theory is just the
mean field chiral Lagrangian theory {\it with BR scaling}.

It is a well-known defect of the mean field Walecka model that the
compressibility modulus $K_0$ is much too high in the model. Now
how does this defect get rectified? The answer must lie in higher loop
corrections going beyond the mean field as the scaling is known to fail
to give the nuclear matter saturation \cite{tjon}. This is also seen in
recent work of Furnstahl et al \cite{tang} who note that by giving
an anomalous dimension 2.7 to the scalar field $\sigma$ with the
Lagrangian suitably implemented with the trace anomaly, they can obtain
the low $K_0\approx 200$ MeV and the suppression of the many-body
terms $\sigma^n$, $n >2$. It is plausible that the anomalous dimension
is mocking up the
quantum loop effects that seem to be needed in the mean field approach given
in \cite{br95-1}. A remarkable observation is that at the anomalous
dimension of $d_a\approx 2.7$, two things happen simultaneously.
One is that the $K_0$ which is large at smaller anomalous dimensions
stabilizes at $\sim 200$ MeV for $d_a\approx 2.7$, stays at that value
for higher $d_a$'s and secondly, it is at this fine-tuned value of $d_a$
that {\it all} multi-body forces get suppressed. This clearly calls
for a simple explanation\cite{chaejun}.

The above result immediately suggests how to calculate kaon-nuclear
interactions in consistency with the nuclear matter properties as given
by Walecka theory. To see this, consider a part of the chiral Lagrangian
that figures importantly in the kaon-nuclear sector
\be
\L_{KN}=\frac{-6i}{8f^2}(\overline{B}\gamma_0 B)\overline{K}\del_t K +
\frac{\Sigma_{KN}}{f^2}(\overline{B}B)\overline{K}K\equiv {\cal L}_\omega
+{\cal L}_\sigma\label{kaonL}
\ee
where $K^T=(K^+ K^0)$. The constant $f$ in (\ref{kaonL}) can be identified
in free space with the pion decay constant $f_\pi$. In medium, however,
it can be modified as we shall see shortly.
In chiral perturbation expansion, the first term corresponds to
${\cal O} (Q)$ and the second term to ${\cal O} (Q^2)$. There is one more
${\cal O} (Q^2)$ term proportional to $\del_t^2$ which will be taken into
account in the numerical results quoted below
but they are not important except for quantitative details.

One can interpret the first term of (\ref{kaonL})
as arising from integrating out the $\omega$
meson as in the baryon sector. The resulting $K^- N$ vector potential in medium
can then be deduced in the same way as for $V_N$:
\be
V_{K^\pm}=\pm\frac{3}{8{f^\star_\pi}^2}\rho.
\ee
Thus in medium, we may set $f\approx f_\pi^\star$ and obtain
\be
V_{K^\pm}=\pm\frac 13 V_N.\label{omegascale}
\ee
This just says that the $\omega$ couples to a {\it matter field} kaon, hence
1/3 of the $\omega$ coupling to the nucleon.  The reason for this matter-field
nature of the kaon is that all nonstrange hadrons become light in dense
medium, so the kaon becomes in some sense heavy. This dual character is known
from the hyperon structure which is well described by considering the kaon
to be heavy as in the Callan-Klebanov model.

As for the second term of (\ref{kaonL}), we use that the kaon behaves as
a massive matter field. We therefore expect that it be coupled to the chiral
scalar $\chi$ as
\be
\L_\sigma =  \frac 13 2 m_K g^\star_\sigma \overline{K}K\chi
\ee
where the factor 1/3 accounts for one non-strange quark in the kaon as compared
with three in the nucleon.
When the $\chi$ field is integrated out as above, we will get,
analogously to the nucleon case,
\be
\L_\sigma = 2m_K \frac 13
\frac{{g^\star_\sigma}^2}{{m^\star_\sigma}^2}\overline{B}B \overline{K}K.
\ee
Comparing with the second term of (\ref{kaonL}), we find
\be
\frac{\Sigma_{KN}}{f^2}\approx 2\frac{m_K}{3}
\frac{{g^\star_\sigma}^2}{{m^\star_\sigma}^2}.\label{relation}
\ee
We can get the $\Sigma_{KN}$ from lattice calculations \cite{fukugita},
$\Sigma_{KN}\approx 3.2 m_\pi$. This gives $f\approx f_\pi^\star$.
Therefore we have
\be
S_{K^\pm}=\frac 13 S_N.
\ee

{\it To summarize: the kaon-nuclear potential gotten from a chiral Lagrangian
and the nucleon-nuclear potential given by Walecka mean field theory
are directly related through BR scaling.}

Given Walecka mean fields for nucleons, we can now calculate the corresponding
mean-field potential for $K^-$-nuclear interactions in symmetric nuclear
matter.
{}From the results obtained above, we have
\be
S_{K^-} +V_{K^-}\approx \frac 13 (S_N-V_N).
\ee
Phenomenology in Walecka mean-field theory gives
$(S_N-V_N)\lsim -600\ {\mbox{MeV}}$ for $\rho=\rho_0$.
This leads to
the prediction that at nuclear matter density
\be
S_{K^-}+V_{K^-}\lsim -200\ {\mbox{MeV}}.
\ee
This seems to be consistent with the result of the analysis in K-mesic atoms
made by  Friedman, Gal and Batty
\cite{friedman} who find attraction at $\rho\approx 0.97\rho_0$ of
\be
S_{K^-}+V_{K^-}=-200\pm 20\ {\mbox{MeV}}.
\ee

An immediate consequence of this mean field description of the kaonic
sector is that kaons will condense in dense neutron star (or nuclear star)
matter at a density
\be
\rho_c\sim 2 \rho
\ee
as found by Lee et al \cite{LBMR} in ChPT to one-loop order.

\begin{figure}
\vskip -7cm
\centerline{\epsfig{file=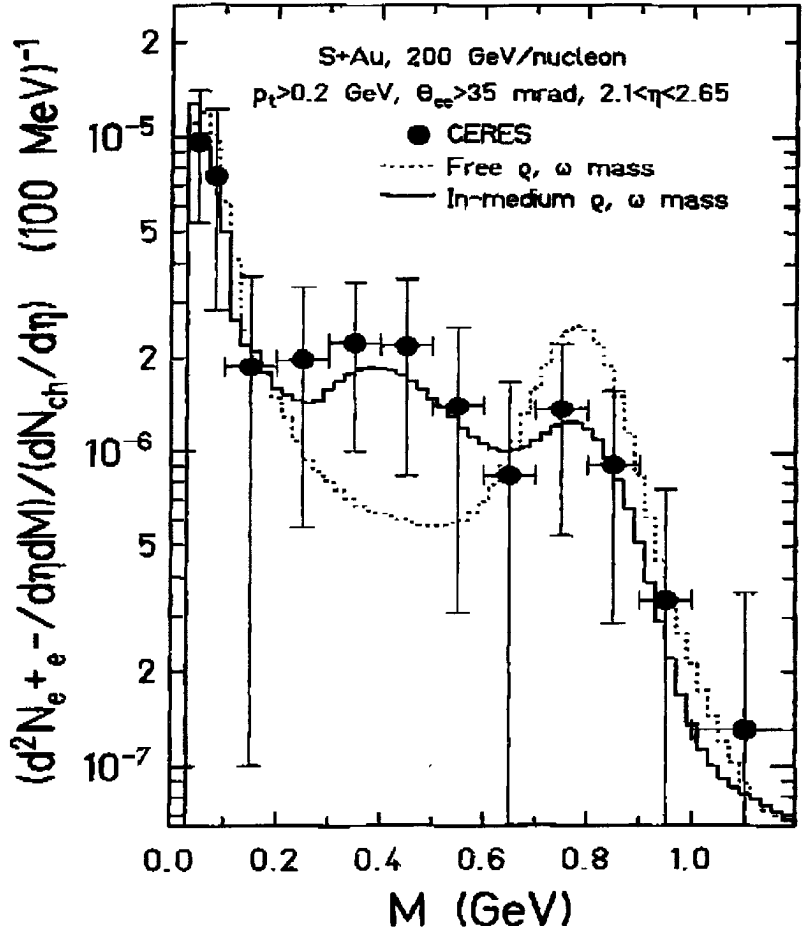}}
\vskip -7cm
\caption[ceres]{The Li-Ko-Brown explanation of the dilepton data of the
CERES collaboration. The dotted line is the theoretical prediction {\it
without}
the scaling of the $\rho$ and $\omega$ mesons and the solid curve
{\it with} the BR scaling. The three lowest mass points are essentially
given by Dalitz pairs, so the relevant data points are the ones for
higher invariant masses. Note that the peak shown at $M\sim 800$ MeV
is predominantly given by
the $\omega$ decay outside of the medium.}\label{ceres}
\end{figure}

To conclude, I make a few remarks on the nature of the scaling
properties.
\ben
\item There are two points to the issue. One is that {\it some or all}
light hadrons may be undergoing a downward mass shift as
density or temperature is
increased. The second is that the scaling is like (\ref{BR}).
These are basically two different issues. To the extent that hadron
masses are generated spontaneously, it is inevitable that at least some
masses should drop. Indeed QCD sum rule calculations do predict the
drop for the vector-meson mass \cite{qcdsum}, the most recent value
being \cite{jin}
\be
m_\rho^\star/m_\rho=0.78\pm 0.08,\ \ \ \rho\approx \rho_0.
\ee

On the other hand, the ``universal scaling" (\ref{BR}), obtained at mean field,
may not be strictly valid. In fact, large $N_c$ arguments imply
$m_N^\star/m_N\approx \sqrt{g_A^\star/g_A} (f_\pi^\star/f_\pi)$. If we look
at the strict $N_c=\infty$ limit, $g_A^\star=g_A$ as we showed in \cite{br91},
the scaling (\ref{BR}) holds but we know that in nuclei, $g_A^\star\approx 1$
and thus in finite nuclei we expect that the nucleon scales somewhat faster
than the pion decay constant, at least up to $\rho\sim \rho_0$.
Whether the ``universal scaling" (\ref{BR}) is consistent
with nature remains to be seen.
\item An intriguing question is how far the scaling (\ref{BR}) can be
pushed in density and/or temperature. Can one use the mean field argument
all the way to the chiral phase transition?

This is a highly relevant question since there are arguments \cite{kocic}
that the second order chiral phase transition relevant to QCD with
2 flavors is of mean field type as in 3D Gross-Neveu model. This suggests that
the mean field chiral theory or Walecka theory could be used to discuss
the QCD chiral phase transition \cite{br95-2}. If correct, this theory
will be very useful for studying phase transitions in heavy ion collisions.
\item There are a large number of experimental projects to measure the
mass shift of hadrons in dense matter, particularly at GSI, CEBAF and
CERN using dilepton-pair production.
Recent data from the CERES collaboration \cite{CERES}
on $e^+ e^-$ pairs in S on Au collisions at 200 GeV/nucleon can be understood
in terms of the scaled mass of the vector mesons in medium, primarily due
to density effect. In a recent paper, Li, Ko and Brown \cite{likobrown}
have shown that the enhancement of the produced lepton pairs observed
in the range of invariant mass $300\ {\mbox{MeV}}\lsim M \lsim 550\
{\mbox{MeV}}$ can be explained simply by the BR scaling in the $\rho$
mass as the pairs are produced mainly through $\pi^+ \pi^-\rightarrow
\rho^\star
\rightarrow e^+ e^-$. The fit is given in Fig.\ref{ceres}.
The analysis is a complex one involving the assumption of an expanding
fireball in chemical equilibrium, but
the economy of the explanation and the quantitative success make it
quite compelling.
\een
\subsection*{Acknowledgments}
\indent

I am very grateful for continuing discussions with Gerry Brown with
whom most of the ideas described here have been developed. I
would also like to thank my young collaborators Chang-Hwan Lee, Kurt
Langfeld and Tae-Sun Park for their valuable help. This paper was
completed at the Center for Theoretical Physics (CTP) of Seoul National
University. I would like to thank Dong-Pil Min and the
members of the CTP for hospitality and support.

\end{document}